\documentclass[twocolumn,showpacs,pre,superscriptaddress]{revtex4}
\usepackage{graphicx}
\begin{document}
\title{Geometry of lipid vesicle adhesion}
\author{R. Capovilla}
\email{capo@fis.cinvestav.mx}
\affiliation{Departamento 
de F\'{\i}sica,
Centro de Investigaci\'on y de Estudios
Avanzados del IPN,
 Apdo. Postal 14-740,07000 M\'exico, DF,
MEXICO}
\author{J. Guven}%
\email{jemal@nuclecu.unam.mx}
\affiliation{%
Instituto de Ciencias Nucleares,
Universidad Nacional Aut\'onoma de M\'exico,
Apdo. Postal 70-543, 04510 M\'exico, DF,
MEXICO}
\date{\today}
\begin{abstract}
The adhesion of a lipid membrane vesicle to a fixed substrate is examined
from a geometrical point of view. This vesicle is described 
by the Helfrich hamiltonian
quadratic in the mean curvature; it interacts by contact with the
substrate, 
with an interaction energy proportional to the area of contact. We
identify the constraints on the geometry at the boundary of the 
shared surface. The result is interpreted in terms of the balance 
of the force normal to this boundary. No assumptions 
are made either on the symmetry of the vesicle 
or on that of the substrate. 
The strong bonding limit as well as the effect of 
curvature asymmetry on the boundary are discussed. 
\end{abstract}


\maketitle

\section{Introduction}

Geometrical models provide a surprisingly 
robust phenomenological description of the equilibria of 
physical membranes
\cite{Nel.Pir.Wei:89,Pel:94,Lip.Sac:95,Bow.Tra:01,Boa:02}. The
hamiltonian which 
describes the membrane is constructed as a sum of geometrical 
scalars; in particular, lipid 
membrane vesicles are well approximated by one 
which is quadratic in the mean curvature\cite{Can:70,Hel:73,Eva:74}. Such
models
can also be 
extended to model adhesion between vesicles
or between a vesicle and a rigid
substrate\cite{Eva:85,Ser.Hel:89,Sei.Lip:90}, processes
which are increasingly relevant to biophysics.
(Two reviews are \cite{Hel:95,Eva:95}.)
In this paper, we examine one important aspect of this problem,  
the geometry of the contact boundary, which, surprisingly, does not 
appear to have been examined in any generality.

To model the interaction one can exploit, as
for an isolated membrane, the geometrical scalars characterizing 
the surface of contact
as well, perhaps, as its boundary. As such, this is not a model
of the adhesion of individual molecules to specific sites on the membrane,
a task which lies beyond the scope of this continuum description.

In its simplest form, which is the one we consider,
the interaction hamiltonian is proportional to (minus) the 
area of contact. The energy associated with
the boundary of the contact region is ignored. 
Axially symmetric configurations  have been studied thoroughly in this `ideal'
context\cite{Sei.Lip:90,Sei.Lip:95} .
In \cite{Sei:91}, the adhesion of `linear' vesicles 
in two dimensions was considered.
More recently, in \cite{Blo.Sag:99,Blo.Sag:99a} and
\cite{Tor.Fou.Gal:01} perturbation theory has been developed in 
the strong bonding limit, in which the 
bending energy itself is small compared to that of adhesion.
Non-axially symmetric configurations of an adhering vesicle under the effect of 
gravity were studied in
\cite{Kra.Sei.Lip:95} using numerical techniques.
We note that a more realistic treatment of adhesion considering
chemically structured or rough surfaces has been provided in \cite{Swa.And:01}.

For definiteness, we will assume that one of the interacting surfaces 
is a fixed substrate, although it is simple to  relax this restriction. 
We will {\it not}, however, assume that this surface is flat. 
More significantly, 
we relax the assumption that the vesicle geometry is axially symmetric.
In previous axially symmetric work the geometric origins of the boundary 
conditions are not clear, because the technique is tailored 
so finely to the symmetry; nor is it clear to what
extent they will survive the relaxation of symmetry. This is a less than desirable 
situation in a model which is intrinsically geometric to begin with.
Of course, one is also interested
in geometries which are not axially symmetric: to mention just one context
where this would be the case,
we note that all configurations with a negative area difference
appear to be inconsistent with axial symmetry \cite{Sve.Zek:89}. Indeed,
it may also be 
energetically favorable for an initially axially symmetric
vesicle to adhere to a substrate in a manner which breaks its original
symmetry.

Our first approach will be to search for minima of the energy.
To do this, we will exploit the geometrical framework 
introduced recently to describe lipid membranes\cite{Cap.Guv:02a}, and
extended to accomodate edge effects in \cite{Cap.Guv.San:02}. The extension to
adhering geometries introduces its own subtleties due to potential discontinuities at
the boundary of contact: the energy is stationary only when 
appropriate constraints on the vesicle geometry are satisfied on this boundary.
Our treatment of the problem  is divided into three parts. 
To establish our bearings,
we begin in Sect. \ref{surf} with a rederivation of the 
Young equation for a liquid droplet where the bonding to the wall 
competes with the surface tension of the drop. In Sect. \ref{lipid}, we
consider lipid vesicles described by the Helfrich hamiltonian.
Discontinuities at the boundary of the contact region are
discussed in Sect. \ref{disc}.
In Sect. \ref{strong},
establishing 
contact with the recent work in this direction, we study the 
strong bonding limit in which the bending energy is 
ignored but any asymmetry between the layers is accounted for.
In this limit, discontinuities at the boundary of the contact region
are not smoothed; a non-vanishing contact angle does not imply a
divergent energy. Finally, in Sect. \ref{full} we consider the general case.
The finiteness of the 
curvature energy necessarily eliminates an angle of contact 
between the vesicle and 
the substrate; stationary energy  completely fixes
the curvature at the boundary.
This generalizes the situation for axially symmetric shapes, where, as 
is well known, the curvature is completely fixed at the
boundary\cite{Sei.Lip:90}: 
the vesicle radius of curvature normal to the line of contact 
is inversely proportional to the square root of the bond strength, 
the tangential radius as well as the potentially non-vanishing
off-diagonal curvature are both completely fixed by the substrate
geometry.
We demonstrate that the boundary condition is not modified by a curvature
asymmetry.
In Sect. \ref{force},
we use the expressions for the internal stress tensor 
in a lipid membrane developed in \cite{Cap.Guv:02a},
to provide  a 
surprisingly simple interpretation of the boundary condition in terms of the
balance of forces at the boundary for a flat substrate. 
We end with some brief conclusions.

\section{Surface tension dominated model}
\label{surf}
It is worthwhile to first review
the adhesion of a drop of liquid of fixed volume onto a surface.
Here the focus is on the competition between the surface tension 
of the liquid and the attraction between the liquid
surface and the substrate.  
The former tends to reduce the surface area of the drop; the 
latter to increase the area of contact.  
The energy is given as a sum of three terms
\begin{equation}
F = \mu A  - w A_{\rm contact}  - p (V-V_0)\,.
\label{eq:F}
\end{equation}
The energy associated with the constant surface tension $\mu$ of the 
drop is proportional to its total surface area $A$;
that associated with adhesion is proportional to (minus) the 
area of contact $A_{\rm contact}$ between the drop and the substrate.
The parameter $w$ is the attractive contact potential.
The third term involving the Lagrange multiplier $p$ 
implements the volume constraint fixing the enclosed drop volume $V$
at the  value $V_0$.

\begin{figure}
\includegraphics[width=2.6in,angle=-90]{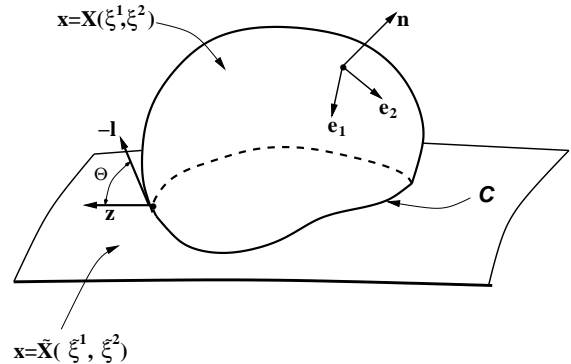}
\caption{\label{fig1}
Definition of the quantities used in the description of the
geometry of adhesion.}
 \end{figure}

The equilibrium drop configurations are those at which  
the  energy (\ref{eq:F}) is stationary. The problem posed here differs from the 
standard isoperimetric problem in that the area at different
positions on the surface gets weighted 
according to whether or not it lies in the contact region,
which itself is determined by the outcome of the variational 
problem. Indeed the contact surface might be weighted negatively. 
For physically realistic parameters, however, an equilibrium is realized.
The energy is always bounded from below.

In equilibrium, the curvature of the drop's surface will 
suffer a discontinuity along the boundary $C$ of the contact region.
We parameterize the embedding of the free surface of the droplet 
in three dimensional space as follows: ${\bf x} = {\bf X}(\xi^a)$,
and the substrate ${\bf x}= \tilde{\bf X}(\tilde\xi^a)$, $a=1,2$.
The energy is a functional both of ${\bf X}$ for 
the free surface and $\tilde{\bf X}$ for the region of contact. They 
coincide on $C$, ${\bf X}=\tilde{\bf X}$. See Fig. 1.
We now recast the first two terms appearing in $F$ as
($A_{\rm free}$ is the area of the free surface)
\begin{eqnarray}
  \mu A  &-& w A_{\rm contact}  
 = \mu A_{\rm free}  + (\mu - w) A_{\rm contact}\nonumber\\
&=& \mu \int_{\rm free} d^2 \xi 
\sqrt{\gamma } + (\mu - w)
\int_{\rm contact} d^2 \tilde\xi\sqrt{\tilde \gamma}\,.
\label{eq:FF}
\end{eqnarray}
Here $\gamma$ is the determinant of the metric 
$\gamma_{ab}$ induced on the free surface given by 
$\gamma_{ab} = {\bf e}_a \cdot {\bf e}_b $,
where ${\bf e}_a = \partial_a {\bf X}$ are vectors tangent
to the surface 
Similar definitions hold for the geometrical quantities, indicated with a
tilde, associated with the substrate.
Note that the boundary $C$ may possess disconnected components. 

To derive the equations describing the equilibrium shape of a 
droplet, let us consider  a variation of the embedding of the free surface, 
${\bf X} \rightarrow {\bf X} + \delta {\bf X}$. 
We let ${\bf n}$ denote the unit normal to the free surface.
We decompose the displacement with respect to the
spatial basis adapted to this surface, $\{ {\bf e}_a ,{\bf n}\}$:
$\delta {\bf X} = \Phi^a {\bf e}_a + \Phi\, {\bf n}$.
We have for the corresponding variation of the induced
metric $
\delta_{X} \gamma_{ab} = 2 K_{ab}\Phi 
+ \nabla_{a}\Phi_{b} + \nabla_b \Phi_a.$
The normal deformation is proportional to the extrinsic curvature tensor,
$K_{ab}= {\bf e}_b\cdot  \partial_a {\bf n}$. The mean extrinsic curvature 
is $K = K_{ab}\gamma^{ab}$. 
The tangential deformation is the Lie derivative of $\gamma_{ab}$ along the 
vector field $\Phi^a$; $\nabla_a$ is the covariant derivative compatible with
$\gamma_{ab}$.

On the boundary $C$, the fixed substrate constrains $\delta {\bf X}$ 
to lie along 
the contact region. We will ignore this for the moment,
treating $\delta{\bf X}$ as though it were unconstrained on $C$. Then
variation of $A_{\rm free}$ gives, 
\begin{equation}
 \delta_X A_{\rm free}
= \int_{\rm free} dA\, K \Phi + 
\int_{C} ds\,  l^a \, \Phi_a\,. 
\label{16}
\end{equation}
Here $ l^a$ is the outward pointing normal to $C$ on the free surface;
$s$ is arclength along $C$. We also have that the variation of the
enclosed volume is
\begin{equation}
\delta_{X} V = \int_{\rm free} dA\, \Phi\,.
\label{eq:delV}
\end{equation}
Remote from $C$ only the normal projection of the variation $\Phi$
plays a role in determining the equilibria of the droplet.
This is generally true regardless of the model.  

For this model, the free 
surface satisfies $\mu K = p$, as follows from the first term in
Eq. (\ref{16}), together with Eq. (\ref{eq:delV}). 
Note that there is no boundary term associated with the normal
deformation $\Phi$. 
This contrasts with the tangential deformation whose 
only net physical effect is to induce a displacement of the boundary. 

The boundary deformation we have described is not free: 
the variation $\delta {\bf X}$ on $C$ is constrained to lie tangent 
to the substrate. Without loss of generality, we can also assume that it 
is normal to the boundary $C$, so that 
\begin{equation}
\delta {\bf X} = \Phi_0 \,
\tilde{\bf z}\,,
\label{eq:Z}
\end{equation} 
where $\tilde{\bf z}$ is the outward unit normal to $C$ on the substrate
(see Fig. 1). 
We then have for the integrand appearing in the boundary term in Eq.(\ref{16}), 
$ l^a\Phi_a = {\bf l} \cdot \delta{\bf  X} = {\bf l}\cdot \tilde{\bf z}\,
\Phi_0$,
where ${\bf l} =  l^a\,{\bf e}_a$ is
 the surface vector $ l^a$ treated as a
spatial vector.
The boundary contribution to the variation of the free surface 
$\delta_X A_{\rm free}$ is then 
\begin{equation}
 \delta_X A_{\rm free}
= \int_{C} ds\, {\bf l}\cdot \tilde{\bf z} \,\Phi_0\,.
\label{eq:bc}
\end{equation}

We now consider the variation of the area of contact $A_{\rm contact}$.
The deformation 
$\delta {\bf X}$  of the free surface will induce a
variation in $A_{\rm contact}$ through the boundary 
that they share,
\begin{equation}
 \delta_X A_{\rm contact} =   
\int_{C} ds\, \Phi_0 \,, 
\label{eq:F2}
\end{equation}
which is a two dimensional analog of Eq.(\ref{eq:delV}). Note that the
substrate
need not be planar.
We can now read off the total boundary contribution to the
variation $\delta_X F$, with $F$ given by Eqs. (\ref{eq:F}), (\ref{eq:FF}).
In equilibrium,
we require that 
\begin{equation}
\int_{C} ds\, 
\big[\mu \, {\bf l}\cdot \tilde{\bf z} +  (\mu- w) \big]\,\Phi_0=0 
\end{equation}
for an arbitrary $\Phi_0$.
We therefore conclude that
\begin{equation}
w = \mu  (1 + {\bf l} \cdot \tilde{\bf z})\,.
\label{eq:BB}
\end{equation}
Defining the contact angle $\Theta$ by $\cos \Theta = -{\bf l} \cdot
\tilde{\bf z}$, this expression reproduces the well known Young equation.

\section{Lipid vesicle adhesion}
\label{lipid}

A lipid membrane is modeled by 
the Helfrich bending energy
\begin{equation}
F_b = \alpha \int dA \; K^2\,.
\label{hel}
\end{equation}
For definiteness, we will focus on either of two variants of the model: 
in both versions the enclosed  volume and the total surface area are fixed; in the
spontaneous curvature model, $K$ is replaced by $K-K_0$ in Eq.(\ref{hel}) where
the constant $K_0$ is the spontaneous curvature; in the bilayer couple model,
the area difference, proportional to the integrated mean curvature
\begin{equation}
M = \int dA\, K\,,
\end{equation} 
is also fixed \cite{Sve.Zek:89}. Thus we construct the constrained energy,
\begin{equation}
F = F_b   - w A_{\rm contact} + \mu (A-A_0) + \beta (M-M_0) - p (V-V_0)\,.
\label{eq:rigid}
\end{equation}
Here $\mu$ is now the Lagrange multiplier associated with the area constraint;
likewise $\beta$ is that associated with the area difference constraint.

\subsection{Discontinuities resolved}
\label{disc}

In the simple model discussed in Sect. \ref{surf} there is no energy
penalty
associated with discontinuities along the contact boundary $C$. 
However, both the intrinsic and extrinsic curvature  will suffer a discontinuity 
along $C$. When the curvature of the vesicle contributes to its energy,
such a discontinuity will generally result in a singularity in the
energy. Because this singular contribution has support on $C$,
it is no longer valid to decompose the energy into two parts, 
$F = F_{\rm free} + F_{\rm contact}$, 

This point is well illustrated by considering an axially symmetric
surface. Cylindrical polar cooordinates $\{\rho,z,\varphi\}$ are
introduced  
on $R^3$; 
constant $\varphi$ curves on the surface are parametrized by arclength $\ell$.
The surface is then described completely once $\rho= R(\ell)$ is specified.
The extrinsic curvature tensor consistent with 
axial symmetry is
\begin{equation}
K_{ab}
= 
\ell_a \ell_b K_\ell  + (\gamma_{ab}-\ell_a \ell_b)K_R\,,
\end{equation}
where the scalars $K_\ell$ and $K_R$ are the two principal curvatures,
and $ \ell^a$ is the outward pointing unit normal to the circle
of fixed $\ell$, $\ell^a=(1,0)$. We identify the scalar curvature ${\cal R} = 
2{\rm det}\, K = 2K_\ell K_R$, and $K = K_\ell + K_R$.
Now let $\theta$ be the angle which the tangent to a curve of fixed $\varphi$
makes with the positive $x$ axis.
The principal curvatures are then given by
 \begin{equation}
K_\ell = \theta'\quad,\quad
K_R ={\sin \theta\over R}\,.
\end{equation}
The prime denotes a derivative with respect to $\ell$.
We have for the integrated mean curvature
\begin{equation}
M = 
2\pi \int d\ell R \, \left(\theta' + {\sin\theta\over R} \right) 
\,.
\end{equation}
Suppose that $\theta$ suffers a 
discontinuity $\Theta$ on the circle at $\ell=\ell_0$, so that
$\theta(\ell)\approx \Theta \,H(\ell-\ell_0)$, where $H$ is the 
step  function. There is a finite contribution from this circle given by
\begin{equation}
\int_C dA \,K
= 
2\pi\int_{\ell_0-\epsilon}^{\ell_0+\epsilon} d\ell\, \theta' R
= 2\pi R (\ell_0)	\Theta \,.
\end{equation}
The mean curvature is thus integrable across the discontinuity.
In general we have the decomposition 
\begin{equation}
M  = \int_{\rm drop/C}
 dA\, K - \int_C ds \, \arccos ({\bf l}\cdot \tilde{\bf z})\,,
\label{eq:kparts}
\end{equation}  
with $\cos \Theta = -{\bf l}\cdot \tilde{\bf z}$, and 
where the notation for the normals is that introduced in Sect.\ref{surf}.

We note that the Gauss-Bonnet invariant for a vesicle
of spherical topology can likewise be decomposed 
\begin{equation}
\label{EH}
\int_{\rm drop /C}
dA\, {\cal R}
+ 2 \int_C ds \, \Delta\kappa = 8\pi\,,
\end{equation}
where $\Delta\kappa = \tilde\kappa - \kappa$ is
the 
discontinuity in the geodesic curvature of $C$.
For an axially symmetric vesicle, the value of the latter term is
$2\pi \cos\Theta$.
(An analogous decomposition of the Hilbert-Einstein
action arises in the study of the dynamics of thin-shells in 
general relativity \cite{Far.Gut.Guv:90}.) As long as the adhering vesicle
remains intact,
the Gauss-Bonnet invariant will not play a role in adhesion. 
Though each of the two components appearing in Eq.(\ref{EH}) will
behave non-trivially under deformation of the surface, their sum will
not change.

The geometric invariant $K^2$ does possess a  singularity at a curvature 
discontinuity. We note that this singularity is identical to that arising from the 
alternative quadratic invariant, $K_{ab}K^{ab}$. This is because the Gauss-Codazzi 
equation, ${\cal R} =
K^2 - K_{ab} K^{ab} $,
identifies their difference as 
the scalar curvature ${\cal R}$ which according to Eq.(\ref{EH})
picks up a finite contribution at a discontinuity. 
In an axially symmetric geometry the troublesome term in 
the bending energy is $\theta'{}^2$: 
 \begin{equation}
\int dA  \,K^2 \approx  
2\pi\int d\ell \,R\,\theta'^2 + \cdots 
\,.
\end{equation}
The $\theta'^2$  term appearing in the integrand gives rise to a 
delta function squared singularity across the boundary. 
To eliminate the corresponding divergence in the energy,
we {\it do} require $\Theta =0$. The surface must be differentiable 
across $C$. It is straightforward to bootstrap this axially 
symmetric analysis to
the general case by introducing Gaussian normal coordinates adapted 
to the boundary. In general, we require that 
$\Theta=0$ or $\tilde{\bf z} = - {\bf l}$.

\subsection{Strong Bonding Limit}
\label{strong}

Before adressing the full problem, let us consider the strong bonding limit,
$\alpha << w A$. At lowest order the bending energy $F_b$
is ignored in Eq. (\ref{eq:rigid}), and the variational problem reduces to
the the minimization of the 
contact energy subject to the three constraints. (In this section, we will
use the language appropriate to the bilayer couple model.)
Whereas $\Theta$ necessarily vanishes on the contact boundary for the 
Helfrich hamiltonian, it need not vanish in this limit.

Let us first consider the variational problem on the 
free surface of the vesicle.
Under a tangential deformation  of this surface any
scalar function
${\cal F}$, and in particular ${\cal F}= \mu + \beta K$,
transforms as a divergence which is transferred to the boundary:
\begin{equation}
\delta_{||} \int_{\rm free} dA\, {\cal F} 
= \int_C ds \, l_a \, \Phi^a \, {\cal F}\,.
\label{eq:bb}
\end{equation}
This is because $\delta_{||} {\cal F} = \Phi^a \partial_a {\cal F}$.
The details  of ${\cal F}$ are irrelevant.
Since $\Phi_a$ is constrained to lie tangent to the contact region, from
Eq. (\ref{eq:Z}), with ${\cal F} = \mu + \beta \, K$, we 
have then
\begin{equation}
\delta_{{||}} \int_{\rm free} dA\, (\mu + \beta \, K ) = 
\int_C ds \, (\mu + \beta K )\, {\bf l}\cdot \tilde{\bf z}\, \Phi_0 \,.
\label{eq:bb1}
\end{equation}
Note that this expression reduces to Eq.(\ref{eq:bc}) when $\mu
=1, \beta=0$.

We now examine a normal deformation of the free surface.
The Euler-Lagrange equation which determines the 
local vesicle equilibrium shape is obtained 
by demanding that the energy be stationary 
with respect to normal deformations of the free surface. 
Its derivation within the present framework 
has been discussed in detail elsewhere \cite{Cap.Guv:02a}. Let us 
focus on the normal deformation of the new ingredient with
respect to the model discussed in Sect. \ref{surf} appearing 
in Eq.(\ref{eq:rigid}) which is $M$. We consider the contributions 
from the free surface, $C$ and 
the contact region as given by Eq.(\ref{eq:kparts}) separately.
We begin with $M_{\rm free}$. We have
\begin{equation}
\delta_{\perp} M_{\rm free} = \int_{\rm free} dA\, {\cal R}\Phi
- \int_C ds  \, \nabla_\perp \Phi \,.
\label{eq:delik}
\end{equation}
We have used $\delta_\perp K = - \nabla^2 \Phi - K^{ab}K_{ab}\Phi$, 
as well as the Gauss-Codazzi equation, and $\nabla_\perp = l^a \nabla_a$ denotes
the derivative
along ${\bf l}$.
It is now straightforward to read off the bulk Euler-Lagrange equation,
\begin{equation}
\mu\, K + \beta\, {\cal R} = p\,.
\end{equation}
Note that this equation is second order in derivatives.

To proceed with the determination of the boundary conditions, we need to 
identify 
the independent unconstrained variation at the interface.
We identify these as $\Phi_0=\tilde{\bf z}\cdot \delta {\bf X}$ and its
derivative along $\tilde{\bf z}$, $\tilde\nabla_\perp \Phi_0$.
We will, however, continue to use
$\nabla_\perp \Phi_0$ to denote
${\bf l}\cdot \tilde{\bf z} \tilde\nabla_\perp\Phi_0$.
We note that the normal deformation at the boundary, using Eq.
(\ref{eq:Z}) is
\begin{equation}
\Phi = {\bf n}\cdot \delta {\bf X}
     = {\bf n}\cdot \tilde{\bf z} \,\Phi_0\,,
\label{eq:ZZ}
\end{equation}
and on the boundary  $C$, its normal derivative is
\begin{eqnarray}
 \nabla_\perp \Phi &=&
{\bf \nabla}_\perp \left({\bf n}\cdot \tilde{\bf z}\,
\Phi_0\right)\nonumber\\
     &=&
({\bf n}\cdot \tilde{\bf z}) \,{\bf \nabla}_\perp \Phi_0
+\Phi_0\; \tilde{\bf z}\cdot \nabla_\perp {\bf n}
+ \Phi_0\; {\bf n} \cdot \nabla_\perp \tilde{\bf z}
\nonumber\\
&=&
({\bf n}\cdot \tilde{\bf z}) \nabla_\perp \Phi_0
+ ({\bf l}\cdot \tilde{\bf z}) [K_\perp + ({\bf l} \cdot \tilde{\bf z})
\tilde{K}_ \perp ]  \Phi_0\,,
\label{eq:ZZZ}
\end{eqnarray}
where we have defined $K_\perp =K_{ab} l^a l^b $, and
$\tilde{K}_{\perp} = \tilde{K}_{ab} \tilde{z}^a \tilde{z}^b$.
We have used the fact that ${\bf n} \cdot \nabla_\perp \tilde{\bf z}=
({\bf l} \cdot \tilde{\bf z})^2 \tilde{K}_{ \perp} $ as well as
$\nabla_\perp {\bf n}=  l^a K_{a}{}^b {\bf e}_b$.
Therefore the boundary contribution to Eq. (\ref{eq:delik}) takes the
form
\begin{equation}
\delta_{\perp} M_{\rm free} =
- \int_C ds  \{({\bf n}\cdot \tilde{\bf z}) \nabla_\perp \Phi_0
+ ({\bf l}\cdot \tilde{\bf z}) [ K_\perp + ({\bf l} \cdot \tilde{\bf z})
\tilde{K}_{\perp }] \Phi_0 \}
\,.
\label{eq:delik1}
\end{equation}

For the boundary contribution $M_C$, we have 
\begin{equation}
\delta_X M_C
= \int_C ds\,\left[ \Theta\; \tilde{\kappa} \Phi_0
+  \delta_X \Theta \right]\,.
\end{equation}
We emphasize that this term contributes {\it only}
to the strong bonding limit.
The first term comes from the variation of arclength: $\delta_X ds =
\tilde{\kappa} \Phi_0$.
We now exploit the fact that
$\cos\Theta = -{\bf l}\cdot \tilde{\bf z}$ to express $\delta_X \Theta $
in terms of $\delta_X ({\bf l}\cdot \tilde{\bf z})$;
to evaluate the latter we note that
$\delta_X \Theta = ( \sin \Theta )^{-1} \delta_X ({\bf l} \cdot
\tilde{\bf z})$, and that
\begin{equation}
\delta_X ({\bf l}\cdot \tilde{\bf z}) = (\delta_X {\bf l})\cdot \tilde{\bf
z}
= {\bf n} \cdot \tilde{\bf z} ( {\bf n} \cdot \delta_X {\bf l} )\,.
\end{equation}
In general, we have for the tangent vectors to the free surface,
\begin{equation}
{\bf n} \cdot \delta_X {\bf e}_a =
\nabla_a \Phi - K_{ab} \Phi^b \,.
\label{dele}
\end{equation}
so that
\begin{equation}
(\delta_X {\bf l})\cdot \tilde{\bf z} = \big(\nabla_\perp\Phi - ({\bf
l}\cdot
\tilde{\bf z})\,K_\perp\Phi_0\big)\,
{\bf n}\cdot \tilde{\bf z}\,.
\label{eq:lz}
\end{equation}
Note that $\delta_X M_C$ need not vanish
even when $M_C$ itself does.

Finally, the induced change in $M$ on the contact region due to the 
displacement of $C$ is just
\begin{equation}
\delta_X M_{\rm contact} = \int_C ds \, \tilde{K} \Phi_0\,.
\end{equation}

It is now straightforward to read off the 
boundary condition by equating the coefficient of 
$\Phi_0$ for the corresponding expression for $F$ to zero:
\begin{equation}
({\bf l}\cdot \tilde{\bf z}) (\mu + \beta K_\parallel)
 + (\mu - w) + \beta \tilde{K} 
+ \beta \Theta\, \tilde\kappa
 =0\,.
\label{eq:sc}
\end{equation}
We have used the fact that $K$ can be expressed as 
$K= K_\parallel + K_\perp$, 
where $K_\parallel= K_{ab}t^a t^b$ is the projection of $K_{ab}$ onto the
unit tangent to $C$, $t^a$.
In general, some simplification is possible by using the identity
\begin{equation}
K_\parallel = {\bf n}\cdot \dot {\bf t} =
\cos\Theta \tilde{K}_{\parallel} + \sin\Theta \tilde\kappa\,.
\end{equation}
In particular, for axially symmetric geometries, we note that
$K_\parallel = K_R$ is consistent with $\tilde{K}_{\parallel}=
\sin\Psi/R$, and
$\tilde\kappa = \cos\Psi/R$, with the identification
$\theta = \Psi+ \Theta$.

Note that Eq. (\ref{eq:sc}) is consistent with Ref.
\cite{Blo.Sag:99} where an 
axially symmetric (indeed spherical) vesicle adhering to a flat substrate 
($\tilde{K} =0$) is described.

\section{ No approximations}
\label{full}

We now examine the general case, as given by the energy
(\ref{eq:rigid}), including the bending energy $F_b$. As discussed in
Sect. \ref{disc}, in order to avoid
discontinuities at the boundary we 
impose $\Theta =0$ or ${\bf l}\cdot \tilde{\bf z}=-1$ as a constraint.

For the tangential deformation  of $F_b$, from Eq. (\ref{eq:bb}),
we have immediately 
\begin{equation}
\delta_\parallel F_b = - \alpha \int_C ds \, K^2 \, \Phi_0\,.
\end{equation}

The novel non-trivial boundary term associated with the normal 
deformation of the free surface originates in the term 
\begin{equation}
2\alpha \int_{\rm free} dA \, K \, \gamma^{ab}
\delta_{\perp} K_{ab} 
\end{equation}
contributing from the variation of $F_{b\; \rm free}$. 
Modulo the free bulk shape equation (described in \cite{Cap.Guv:02a}),
there remains
\begin{equation}
\delta_{\perp} F_{b \; \rm free} =
2\alpha \int_C ds  \left( \Phi \, \nabla_\perp K - K \, \nabla_\perp \Phi
\right)\,.
\end{equation} 
We note that $\Phi ={\bf n}\cdot \tilde{\bf z} \,\Phi_0 =0$, so that the
first
term vanishes. For the second term, on $C$, Eq.(\ref{eq:ZZZ}) gives 
$
\nabla_\perp \Phi = ( \tilde{K}_{ \perp}
- K_\perp ) \Phi_0
$, so that the novel contribution is
\begin{equation}
\delta_{\perp} F_{b \; \rm free} =
2\alpha \int_C ds  K \, ( K_\perp  - \tilde{K}_{\perp}
)\,
\Phi_0\,.
\end{equation}

It is now straightforward to read off the 
boundary condition (there is no term proportional to $\nabla_\perp\Phi_0$)
\begin{equation}
- \alpha K^2  - \beta K_\parallel + \alpha \tilde{K}^2
+ \beta \tilde{K}_{\parallel} + 2 \alpha
( K_\perp - \tilde{K}_{\perp} ) K  =w\,.
\label{eq:ksq}
\end{equation}
If we now use the fact that $K_\parallel-\tilde{K}_{
\parallel}=0$
when $\Theta=0$, Eq.(38) reduces to the remarkably simple expression
\begin{equation}
K_\perp -  \tilde{K}_{\perp}
=\sqrt{w/\alpha}\,.
\label{eq:ksqr}
\end{equation}
This expression is independent of $\beta$.

The curvature $K_\perp$ is completely fixed at the
boundary by Eq.(\ref{eq:ksqr}).
We note that the off-diagonal term with respect to the basis $l^a$ and
$t^a$, $K_{\parallel\perp}= l^a t^b K_{ab}= {\bf n}
\cdot \dot {\bf l}$ will not generally be zero. Just like
$K_\parallel$, however, it will is completely determined by its
substrate counterpart,
$K_{\parallel\perp} = \tilde{K}_{\parallel\perp}$.
Thus, all three components
of the curvature are fixed at the boundary.

If the substrate is flat at the boundary, we have $K_\perp=
\sqrt{w/\alpha}$.
For an axially symmetric shape, $K_\perp = K_{\ell}=\theta'$ and
Eq.(\ref{eq:ksqr}) therefore reproduces the well known
boundary condition \cite{Sei.Lip:90},
\begin{equation}
\theta'= \sqrt{w/\alpha}\,.
\label{eq:axad}
\end{equation}
We note that
if the substrate is axially symmetric and not flat, Eq. (\ref{eq:axad})
is modified to
\begin{equation}
\theta'- \Psi'= \sqrt{w/\alpha}\,,
\end{equation}
where $\Psi'$ is the curvature along a meridian of the substrate.
This agrees with
the expression given in footnote 14 of \cite{Sei.Lip:90}.

\section{Stresses at the boundary} \label{force}

In equilibrium, the forces directed along the normal from the boundary
into the membrane must balance. In \cite{Cap.Guv:02a}, it is shown that
the stress tensor for the model (\ref{eq:rigid})  can be expressed as
\begin{eqnarray}
{\bf f}^a &=& \big[\alpha K (2 K^{ab} - K \gamma^{ab}) +
\beta (K^{ab} - K \gamma^{ab}) \nonumber \\ &-& \mu   
\gamma^{ab} \big] {\bf e}_b - 2 \alpha \nabla^a K {\bf n}\,.
\end{eqnarray} This is the stress tensor on the free surface. It satisfies
\begin{equation} \nabla_a\,{\bf f}^a = p\, {\bf n} \label{cons}
\end{equation} at each point.

Let us for simplicity suppose that the substrate is flat, so that
$\tilde{K}_{ab} = 0$. The
corresponding
stress tensor in that part of the vesicle which is bound to the 
substrate $\tilde{\bf f}^a$ is then 
\begin{equation}
\tilde{\bf f}^a = (w
-\mu )\,  \tilde\gamma^{ab} \tilde{\bf e}_b \,,
\end{equation} 
which is isotropic.
Note that $\tilde{\bf f}$ does not satisfy the  
conservation law Eq.(\ref{cons}). Thus the construction of a Gaussian 
pillbox of infinitesimal thickness on the boundary does not lead 
to a useful identity.
We note, however,  that ${\bf l}\cdot{\bf f}^a l_a$ is 
the pressure acting on the boundary 
due to unbalanced stresses in the free bulk at its boundary. We have
\begin{equation}
{\bf l}\cdot {\bf f}^a l_a  =
\alpha K (2 K_\perp - K) - \beta K_\parallel - \mu
 \,.\label{fb}
\end{equation}
Similarly, 
\begin{equation}
\tilde{\bf z}\cdot \tilde{\bf f}^a \tilde{z}_a = w-\mu\,.\label{fc}
\end{equation}
The stresses must balance in equilibrium.
When they do, Eq.(\ref{eq:ksqr}) is reproduced.
This derivation is not only more efficient than the 
variational argument, it also homes in immediately 
on the physics at the boundary.

\section{Conclusions}

We have shown how a combination of simple geometrical and variational 
techniques as well as conservation laws can be applied to 
study the adhesion of vesicles described by a 
geometrical hamiltonian. This provides a useful 
platform for either a numerical or perturbative approach to adhesion,
particularly when 
one is interested in non-axially symmetric shapes. 
Axially symmetric shapes are very special ones.

These techniques also generalize to so-called floppy or egg-carton membranes 
where a term penalizing curvature gradients also appears in the 
hamiltonian \cite{Hel:95,Goe.Hel:96}. Now, not only is the contact
angle 
fixed, but its first derivative vanishes. It is the second derivative which will be 
proportional to the bond strength. The interesting shapes 
are also certainly not axially symmetric. 

\vspace{1cm}

{\sl Note added in proof}: After this work was completed, Ref.
\cite{Ros.Vir:98} was brought to our attention, where  
the adhesion of `linear' vesicles 
in two dimensions is considered.

\begin{acknowledgments}
RC is partially supported by CONACyT under grant 32187-E.
JG is partially supported by CONACyT 
under grant 32307-E and DGAPA at UNAM under grant IN119799.
We thank B. Bozic, S. Svetina, and B. Zeks for useful comments, 
for
bringing
Ref. \cite{Ros.Vir:98} to our attention,  
and for their help in pointing out a mistake in an earlier version of this
paper.
\end{acknowledgments}

%\bibliographystyle{apsrev}
%\bibliography{strings}

\end{document}